\documentclass[preprintnumbers,secnumarabic,superscriptaddress]{revtex4}

\usepackage{graphicx}
\usepackage{amsmath,amssymb,mathrsfs}
\usepackage{fancyhdr}
\usepackage{psfrag}
\usepackage{array,bbm}
\usepackage{rotating}
\usepackage{multirow}
\usepackage{tabularx} 
\usepackage{dcolumn}

\newcommand*{\trans}{\mathrm{T}}                     
\newcommand*{\unitmatrix}{\mathbbm{1}}

\newcommand*{\tvec}[1]{\ensuremath{\boldsymbol{\mathrm{#1}}}}           

\newcommand*{\eweakgroup}{\mbox{$SU(2)_L \times U(1)_Y$} }
\newcommand*{\emgroup}{\mbox{$U(1)_{em}$} }

\newcolumntype{.}{D{.}{.}{-1}}
\DeclareMathOperator{\tr}{tr}
\DeclareMathOperator{\re}{Re}
\DeclareMathOperator{\im}{Im}

\begin{document}


\title{On exact minimization of Higgs potentials}

\author{Markos Maniatis}
\affiliation{Departamento de Ciencias B\'a{}sicas, Facultad de Ciencias,
		Universidad del B\'i{}o-B\'i{}o, Chill\'a{}n, Chile}
\author{Dhagash Mehta}
\affiliation{
Dept of Mathematics, North Carolina State University, Raleigh, NC 27695-8205, USA}




\begin{abstract}
Minimizing the Higgs potential is an essential task in any
model involving Higgs bosons. Exact minimization methods proposed
in the literature are based on the polynomial form of the potential.
These methods will in general no longer work if loop contributions to the potential
are taken into account. We present a method to keep the
tree level global minimum unchanged in passing to the 
effective potential. We illustrate
the method for the case of the Minimal Supersymmetric Model~(MSSM).
\end{abstract}

\maketitle

\section{Introduction}
\label{s:introduction}

The search for the Higgs boson at the Large Hadron Collider
has initiated many theoretical studies of Higgs physics --
see for
instance~\cite{Santos:2010zz,Ma:2011ea,Ferreira:2011aa,Mader:2012pm,
Maniatis:2009vp,Maniatis:2009by,Brehmer:2012hh,Espinosa:2012ir,Arbey:2013fqa,Krawczyk:2013gia,Bechtle:2013xfa}.
The detection of the minima of a Higgs potential of
any model is essential in order to determine the particle spectrum as well as
the electroweak symmetry--breaking behavior.
However, finding the minima of a Higgs potential is in general a non-trivial task.
For instance, in the general two-Higgs doublet model~(THDM),
already at tree level the Higgs potential
consists of three bilinear terms as well as seven quartic terms in
the Higgs-boson doublet fields; for a recent review of the THDM see
for instance~\cite{Branco:2011iw}.
In contrast, due to the restrictions imposed by gauge invariance and renormalizability, 
in the Standard Model with one Higgs-boson doublet, there
is only one bilinear and one quartic term possible.

Even though the potential is in general non-linear and complicated, an exact solution
is often accessible if the potential is of the polynomial form.
All stationary points follow immediately from the solutions of the
gradient equations of the potential yielding systems of equations.
In particular, assuming that the potential is bounded from below,
the global minima are the stationary points with the lowest potential
value.
It was shown for the general
THDM, as well as for even more involved cases, that exact solutions
can be achieved by employing the Gr\"obner basis approach
or homotopy continuation~\cite{Maniatis:2006jd,Maniatis:2012ex,Mehta:2009,Mehta:2009zv,Mehta:2011xs}.
It is found that the structure of minima 
of models beyond the Standard Model Higgs sector
is in general
very rich. In particular, different stationary points
lie often far away in terms of the Higgs-boson fields but nevertheless
correspond to very close potential values. For instance
in a study
of a supersymmetric model with two Higgs-boson doublets
and one Higgs-boson singlet~\cite{Maniatis:2006jd} (the
model studied is the NMSSM -- for reviews see~\cite{Maniatis:2009re,Ellwanger:2009dp}),
nearly degenerate vacua were detected for very different
points in the Higgs-boson fields. This finding
makes it questionable if even at tree-level the conventional numerical methods
will certainly find the correct
global minima for potentials with multiple Higgs bosons.

The situation is even more complicated if quantum corrections are
taken into account. In general, all massive particles of a model
will contribute to the Higgs potential via loops.
Moreover, these loop contributions will in general introduce
transcendental, that is non-polynomial, functions into the Higgs potential.
The appearance of transcendental functions makes it impossible
to apply approaches like the Gr\"obner-basis approach or
homotopy continuation directly, since these approaches work only
with polynomials.

Nevertheless, in the following we will demonstrate that
there is a way to preserve the global minimum.
The idea is to fix the renormalization scale, which inevitably
appears in the loop contributions, such that the gradient
of the one-loop contributions to the potential vanishes at the vacuum.
With this assignment of the renormalization scale, the tree-level
global minimum is unchanged if we consider the full one loop effective
potential. We will illustrate the procedure for the
case of the Minimal Supersymmetric Standard Model~(MSSM)
in detail in the following;
for a review of the MSSM see for instance~\cite{Martin:1997ns}.
We note that the approach outlined here
should be applicable to other Higgs potentials
in an analogous way.

Let us eventually mention some examples of studies
of the minimization of Higgs potentials:
the approaches~\cite{Maniatis:2006jd,Maniatis:2012ex,Mehta:2011xs}
are devoted to the minimization of the tree-level potential of
multi-Higgs models. In these approaches, the stationarity
equations are systematically solved employing the Gr\"{o}bner-basis
approach or homotopy continuation.
In the approaches~\cite{Battye:2011jj,Degee:2012sk}
the vacuum is deduced in highly symmetric potentials.

In~\cite{Gamberini:1989jw} the 
renormalization-group improved tree-level Higgs potential
of the MSSM is studied critically: a comparison with
the effective one-loop potential shows that the
loop corrections can be large and have to be taken
into account in general. In particular it is shown
that the vacuum-expectation values of the two Higgs
bosons depend strongly on the choosen renormalization scale.
This situation is improved
drastically when the full one-loop potential is considered.
Based on this finding, in~\cite{de Carlos:1993yy} the minimization
of the effective MSSM Higgs potential in context with fine tuning is studied. 
Is is pointed
out that the loop contributions change the structure
of the stationary points substantially and in general all contributions
to the effective potential have to be taken into account.
The work~\cite{Casas:1995pd} is also devoted to the study
of the stationary solutions of the effective potential.
In this work the renormalization scale is chosen such that one of
the vacuum-expectation values of the two Higgs-boson doublets
is shared between tree-level and effective potential. 
In the work~\cite{CamargoMolina:2012hv} the framework of homotopy continuation
is applied to minimize the tree-level potential. Starting
from the tree-level stationary points, numerical methods are applied
to determine the minima of the effective loop potential. 

\section{Minimization method}
\label{s:MSSM}

In this section we illustrate the minimization method for the case
of the MSSM. Let us start with the tree-level Higgs potential.
In the MSSM we get contributions to the tree level Higgs potential
from $F$ and $D$ terms, as well as from the soft
supersymmetry-breaking terms. Explicitly, we have~\cite{Martin:1997ns}
\begin{equation}
\label{MSSMpot}
\begin{split}
&V_F = |\mu|^2 (\varphi_1^\dagger \varphi_1+\varphi_2^\dagger \varphi_2) ,\\
&V_D = \frac{g_1^2 + g_2^2}{8}(\varphi_1^\dagger \varphi_1- \varphi_2^\dagger
\varphi_2)^2
+\frac{g_2^2}{2} \big( (\varphi_1^\dagger \varphi_1)(\varphi_2^\dagger \varphi_2)-(\varphi_1^\dagger \varphi_2)
(\varphi_2^\dagger \varphi_1) \big),\\
&V_{\text{soft}} = m_{H_1}^2 (\varphi_1^\dagger \varphi_1) + m_{H_2}^2
(\varphi_2^\dagger \varphi_2)
- (m_3^2 (\varphi_1^\dagger \varphi_2) + H.c.) ,
\end{split}
\end{equation}
with the tree-level potential
\begin{equation}
\label{Vtree}
V^{\text{tree}}= V_F + V_D + V_{\text{soft}}\;.
\end{equation}
In this convention, both Higgs-boson doublets $\varphi_{1/2}$ have the same hypercharge,
$Y(\varphi_{1/2})=+1/2$, but can easily translated to the
more common convention via
\begin{equation}
H_1 =  \epsilon \varphi_1^* \;, \quad
H_2 = \varphi_2 \;, \qquad \text{with}  \quad
\epsilon =
\begin{pmatrix}
\phantom{+}0 & 1 \\ -1 & 0
\end{pmatrix}
\end{equation}
where the doublets carry hypercharges $Y(H_1) = -1/2$, $Y(H_2) = +1/2$.

Let us now consider the dominant loop contributions for the third
family of the quark--squark supermultiplet to the Higgs potential.
Their contribution to the effective one-loop Higgs potential,
\begin{equation}
\label{effpot}
V = V^{\text{tree}} + V^{\text{1-loop}}\;,
\end{equation}
reads~\cite{Coleman:1973jx}
\begin{equation}
\label{VMSSM1}
V^{\text{1-loop}} =
\frac{3}{32 \pi^2} \bigg\{
\tr \bigg[ M_{\tilde{q}}^4 \bigg( \ln\bigg(\frac{M^2_{\tilde{q}}}{Q^2} \bigg) - \frac{3}{2} \bigg)  \bigg]
-
2 \; \tr \bigg[ M_q^4 \bigg( \ln \bigg(\frac{M^2_q}{Q^2} \bigg) - \frac{3}{2} \bigg) \bigg]
\bigg\}.
\end{equation}
Here, $M^2_q$ and $M^2_{\tilde{q}}$ are the squared field-dependent mass matrices
for the quarks and squarks, respectively.
Note that, quite naturally in a supersymmetric theory,
fermions and bosons appear with different sign
in the generic Coleman--Weinberg corrections.
The matrices have the explicit form; see also~\cite{Carena:2000yi}
\begin{equation}
M_q^2 =
\begin{pmatrix}
|h_t|^2 (\varphi_2^\dagger \varphi_2) & h_b^* h_t (\varphi_1^\trans i \sigma_2 \varphi_2) \\
h_b h_t^* (\varphi_1^\trans i \sigma_2 \varphi_2)^* & |h_b|^2 (\varphi_1^\dagger \varphi_1)
\end{pmatrix}
\end{equation}
in the basis $(t, b)^\trans$ and
\begin{equation}
M_{\tilde{q}}^2 =
\begin{pmatrix}
	\begin{pmatrix}\begin{tabular}{m{24pt} m{24pt}} \multicolumn{2}{c}{\multirow{-1}*{$m_{11}$}} \end{tabular}\end{pmatrix}_{\mathbbm{2}\times\mathbbm{2}}&
	\begin{pmatrix}\begin{tabular}{m{12pt} m{24pt}} \multirow{-1}*{$\!\!m_{12}\!\!\!\!\!$} \end{tabular}\end{pmatrix}_{\mathbbm{2}\times\mathbbm{1}}&
	\begin{pmatrix}\begin{tabular}{m{12pt} m{24pt}} \multirow{-1}*{$\!\!m_{13}\!\!\!\!\!$} \end{tabular}\end{pmatrix}_{\mathbbm{2}\times\mathbbm{1}}\\
	\begin{tabular}{m{24pt} m{12pt}} \multicolumn{2}{c}{$(\;m_{21}\;)_{\mathbbm{1}\times\mathbbm{2}}$} \end{tabular}&
	m_{22}&
	m_{23}\\
	\begin{tabular}{m{24pt} m{12pt}} \multicolumn{2}{c}{$(\;m_{31}\;)_{\mathbbm{1}\times\mathbbm{2}}$} \end{tabular}&
	m_{32}&
	m_{33}
	\end{pmatrix}
\end{equation}
in the basis $( (\tilde{t}_L, \tilde{b}_R), \tilde{t}^*_R, \tilde{b}^*_R )^\trans$ with
\begin{equation}
\label{msqent}
\begin{split}
&
m_{11} =m_{\tilde{q}3L}^2 \unitmatrix_2
+ (|h_b|^2 - \frac{g_2}{2} ) \varphi_1 \varphi_1^\dagger
-(|h_t|^2 - \frac{g_2}{2} ) \varphi_2 \varphi_2^\dagger
+(\frac{g_2^2}{4} - \frac{g_1^2}{12}) (\varphi_1^\dagger \varphi_1) \unitmatrix_2
+(|h_t|^2-\frac{g_2^2}{4} + \frac{g_1^2}{12}) (\varphi_2^\dagger \varphi_2) \unitmatrix_2,
\\
&
m_{21} = h_t \mu^* \varphi_1^\trans i \sigma_2 - h_t A_t \varphi_2^\trans i \sigma_2,
\\
&
m_{31} = h_b A_b \varphi_1^\dagger - h_b \mu^* \varphi_2^\dagger,
\\
&
m_{22} = m_{\tilde{t}R}^2 +|h_t|^2 (\varphi_2^\dagger \varphi_2) +
\frac{g_1^2}{3} ( (\varphi_1^\dagger \varphi_1) - (\varphi_2^\dagger \varphi_2)),
\\
&
m_{33} = m_{\tilde{b}R}^2 +|h_b|^2 (\varphi_1^\dagger \varphi_1) -
\frac{g_1^2}{6} ( (\varphi_1^\dagger \varphi_1) - (\varphi_2^\dagger \varphi_2)),
\\
&
m_{23} = h_t h_b^* (\varphi_1^\trans i \sigma_2 \varphi_2) \;.
\end{split}
\end{equation}
Here $g_1$ and $g_2$ are the usual $U(1)_Y$ and $SU(2)_L$ gauge couplings;
$\mu$ is the mass parameter in the superpotential; $h_b$ and $h_t$ are
the bottom and top Yukawa couplings, respectively; $A_b$ and $A_t$ are the
trilinear couplings; and $\sigma_2$ is the second Pauli matrix.

Further supermulitplets which couple to
the Higgs bosons may be
taken into account in a completely analogous way. For simplicity, here we consider only the
dominant contributions of the third quark, respectively squark family.

The task now is to find the minima of the effective potential~\eqref{effpot}.
All stationary points of the potential follow from vanishing gradients
of the potential. Assuming that the potential is bounded from below, the
stationary points with the lowest potential value are the global minima.
Considering the tree-level potential~\eqref{Vtree} the gradient equation
yields systems of polynomial equations. These systems of equations
are in general solvable employing the Gr\"obner-basis
approach or homotopy continuation~\cite{Maniatis:2006jd,Maniatis:2012ex,Mehta:2011xs}.
However, taking loop contributions to the potential into account,
the gradient equations are no longer of the polynomial form; prohibiting
the application of the Gr\"obner or homotopy continuation approaches.
The idea is now to fix the renormalization scale~$Q$ such that
the gradient of the loop contributions vanishes at the vacuum,
\begin{equation}
\label{vanishgradient}
\langle \nabla V^\text{1-loop}(\varphi_1, \varphi_2) \rangle =0\;,
\end{equation}
where the gradient is to be built with respect to the Higgs-boson field
degrees of freedom.
In this way the global minimum is kept unchanged
passing from the tree-level potential to 
the effective potential.\\

Employing the condition~\eqref{vanishgradient} we ensure that the
stationarity solutions of the tree-level potential at the vacuum are preserved.
In particular, stability and electroweak symmetry breaking are
kept unchanged at the vacuum but only the physically irrelevant potential value itself
is changed in general.

In our calculations we will represent the
potential in the bilinear
formalism~\cite{Nagel:2004sw,Maniatis:2006fs,Nishi:2006tg}. Let us briefly
review this approach here:
in the Higgs potential the two Higgs-boson doublets
$\varphi_1$ and $\varphi_2$ appear
as scalar products $\varphi_i^\dagger \varphi_j$ with $i,j \in \{1,2\}$
(this is also the case for the traces in the loop contributions).
These scalar products may be replaced
by the bilinears $K_0$, $K_1$, $K_2$
and $K_3$; for details we refer to~\cite{Maniatis:2006fs},
\begin{equation}
\begin{alignedat}{2}
\label{eq-phik}
\varphi_1^{\dagger}\varphi_1 &= (K_0 + K_3)/2, \quad
\varphi_1^{\dagger}\varphi_2 &= (K_1 + i K_2)/2, \\
\varphi_2^{\dagger}\varphi_2 &= (K_0 - K_3)/2, \quad
\varphi_2^{\dagger}\varphi_1 &= (K_1 - i K_2)/2.
\end{alignedat}
\end{equation}
The bilinears have to fulfill the conditions
\begin{equation}
\label{eq-kconditions}
K_0 \ge 0, \quad K_0^2-K_1^2-K_2^2-K_3^2 \ge 0.
\end{equation}
One of the advantages of the bilinears is that the \eweakgroup gauge degrees are
eliminated. Moreover,
the degree of the potential is also reduced.
With respect to the tree-level potential for the MSSM
in terms of the $K_\alpha$, $\alpha = 0,...,3$, we find
(with the convention of summation over repeated indices)
\begin{equation}
\label{VK}
V^\text{tree} = \xi_{\alpha}  K_\alpha + \eta_{\alpha \beta} K_\alpha K_\beta
\end{equation}
with
\begin{equation}
\label{etaMSSM}
\xi =
\begin{pmatrix}
\frac{1}{2} (m_{H_1}^2+m_{H_2}^2) + |\mu|^2\\
- \re{m_3^2}\\
\phantom{+} \im{m_3^2}\\
\frac{1}{2} (m_{H_1}^2-m_{H_2}^2)
\end{pmatrix} , \qquad
\eta = \frac{1}{8}
\begin{pmatrix}
g_2^2 & 0 & 0 & 0\\
0 & -g_2^2 & 0 & 0\\
0 & 0 & -g_2^2 & 0\\
0 & 0 & 0 & g_1^2
\end{pmatrix} .
\end{equation}

Stability and a non-trivial electroweak symmetry breaking of the
tree-level potential require~\cite{Maniatis:2006fs}
\begin{equation}
\label{treestability}
\xi_0 - \sqrt{\xi_1^2+\xi_2^2}>0 \qquad \text{and }
\xi_0 < |\tvec{\xi}|\;,
\end{equation}
with $\xi_0$ and $\tvec{\xi}$ the time-like respectively spatial
components of the four vector $\xi$.
The stationarity conditions of
the Higgs potential
in terms of $K_\alpha$ with respect to the
electroweak symmetry-breaking behavior
can now be easily formulated~\cite{Maniatis:2006fs,Maniatis:2006jd}:

\begin{itemize}
\item Unbroken electroweak symmetry: stationary points of the potential with an unbroken electorweak gauge symmetry are
given by
\begin{equation}
K_0=K_1=K_2=K_3=0.
\end{equation}

\item Fully broken electroweak symmetry: these stationary points are points with
\begin{equation}\label{eq-kcondfull}
K_0>0,\qquad K_0^2-K_1^2-K_2^2-K_3^2 > 0.
\end{equation}
A global minimum of this type has non-vanishing
vacuum-expectation values for the
charged components of the doublet fields,
thus gives fully broken \mbox{\eweakgroup;} see~\cite{Maniatis:2006fs}.
The stationary points of this type are found by requiring a vanishing gradient
with respect to all bilinears $K_\alpha$:
\begin{equation}
\label{eq-stationarityF}
\nabla \; V(K_0, K_1, K_2, K_3) =0.
\end{equation}
Only real solutions of~\eqref{eq-stationarityF} 
fulfilling~\eqref{eq-kcondfull} are valid.

\item Partially broken electroweak symmetry: stationary points corresponding to a partially broken electroweak gauge symmetry
\eweakgroup down to~\emgroup
are points with
\begin{equation}\label{eq-kcondpartial}
K_0>0,\qquad
K_0^2-K_1^2-K_2^2-K_3^2 = 0.
\end{equation}
Employing one Lagrange multiplier $u$, these stationary points are given
by the real solutions of the system of equations
\begin{equation}
\label{eq-stationarityP}
\nabla
\big[
V(K_0, K_1, K_2, K_3)
- u \cdot (K_0^2-K_1^2-K_2^2-K_3^2)\; \big] = 0, \qquad
K_0^2-K_1^2-K_2^2-K_3^2 = 0,
\end{equation}
Only real solutions with $K_0>0$ are valid solutions of this type.
\end{itemize}

As long as the potential is bounded from below,
the stationary solutions with the lowest potential value
are the global minima. Evidently, only the global minima
with the partially broken electroweak symmetry are in general
acceptable. Let us note that a metastable minimum,
that is,
a minimum with the desired
electroweak symmetry breaking which is {\em not} the global
minimum, may have interesting
phenomenological consequences. For a recent study
of metastable vacua in the THDM we refer to~\cite{Barroso:2013awa}.

Finally, let us give the condition for a vanishing gradient of
the loop contribution to the potential~\eqref{vanishgradient}
in terms of bilinears,
\begin{equation}
\label{vanishgradientK}
\langle \nabla V^\text{1-loop}(K_0, K_1, K_2, K_3) \rangle =0\;,
\end{equation}
that is, the gradient has to be built with respect to the four bilinears
$K_\alpha$, $\alpha=0,...,3$.
Employing the condition~\eqref{vanishgradientK} we have to express the effective loop part 
of the potential~\eqref{effpot}--\eqref{msqent} in terms of bilinears.


\section{Results}
\label{s:numerics}

In this section we demonstrate the detection of the
minima and illustrate our method.
Guided by one of the SoftSusy~\cite{Allanach:2001kg} benchmark sets,
called  ``cmssm10.1.1.spec'' (this is a {\em constrained} case),
we choose as an example the parameters as given in Table \ref{parameters}.
\begin{table}[ht]
\begin{tabular}{ll}
$g_1=0.362414$, $g_2=0.643026$ & $U(1)_Y$ resp. $SU(2)_L$ gauge couplings\\
$m_{H^\pm}= 722.471$~GeV & charged Higgs-boson masses\\
$m_{A^0} = 718.195$~GeV & CP odd Higgs-boson mass\\
$\mu = 629.936$~GeV & superpotential Higgs parameter\\
$h_b = 0.135322$, $h_t = 0.861186$ & bottom resp. top Yukawa coupling\\
$A_b = -1304.2$~GeV, $A_t = -880.158$~GeV  & trilinear bottom and top parameters\\
$v = 244.112$~GeV, $\tan(\beta)=9.67352$ & vacuum-expectation value and ratio of vevs $v_2/v_1$\\
$m_{H_1}^2=117504 \text{ GeV}^2$, $m_{H_2}^2= -395338 \text{ GeV}^2$ & soft Higgs-boson parameters\\
$M_{\tilde{b}}=970.737 \text{ GeV}$ & soft scalar bottom parameters\\
\end{tabular}
\caption{\label{parameters} Chosen parameters in our example in case of the MSSM. For details
we refer to~\cite{Allanach:2001kg}.}
\end{table}
The convention for the vacuum-expectation values is
$\langle H_i^0 \rangle = 1/\sqrt{2} v_i$ with $v=\sqrt{v_1^2+v_2^2}$ and
$\tan(\beta)=v_2/v_1$.

Given the set of parameters we start with a study of the tree-level potential. As pointed
out before, we employ the bilinear formalism with a potential as given in~\eqref{VK}.
The parameters in terms of the bilinear formalism follow directly
from~\eqref{etaMSSM}.
We check that the conditions for stability and
a non-trivial minimum, that is~\eqref{treestability}, are fulfilled.

We proceed with a systematic search for the stationary points as
outlined in the last section.
Explicitly, we employ the Gr\"obner-basis approach to solve all
stationarity equations (for a brief introduction to the Gr\"obner-basis
approach we refer to the appendix in~\cite{Maniatis:2006jd}.)
We use the Gr\"obner-basis approach as implemented
in the Singular program package which is freely available online~\cite{Singular}.
In general there are two complex solutions, where, with the parameters chosen,
both turn out to be real and pass the condition $K_0 \ge 0$.
The two stationary points are given in Table \ref{treemin}. 
Obviously, for vanishing
fields $\varphi_1$ and $\varphi_2$, corresponding to vanishing bilinears $K_0,...,K_3$
we have a stationary point with an unbroken electroweak symmetry. 
From~\eqref{Vtree} or \eqref{VK} we see that this stationary point has a vanishing potential value.
Furthermore, there is one stationary point with the lowest potential value
which has the right, that is, partially broken, electroweak-symmetry breaking.
Since the potential is bounded from below and there is no deeper stationary point, this
stationary point is the global minimum, or vacuum, of the tree level potential.
The vacuum-expectation value and the ratio of the two Higgs-boson doublets~($\tan(\beta)$)
follow from
$K_0=1/2 v^2$, $K_1= v^2 \sin(\beta) \cos(\beta)$,
$K_2 = 0$, $K_3 = 1/2 v^2 \cos(2 \beta)$ at this point.

We ensure that the tadpole conditions at the vacuum are fulfilled -
actually the parameters $m_{H_1}^2$, 
$m_{H_2}^2$, $m_{A^0}$, as given in Tab.~\ref{parameters}, are fixed such that
the tree-level tadpole conditions vanish at the desired vacuum. 
As a consistency check we verify that the bordered Hessian 
corresponds to a local minimum at the vacuum. \\

With the preceding preparations we now approach the effective potential~\eqref{effpot}. 
As outlined above, in order to keep the global minimum passing to
the effective potential, we
require that the gradient of the one-loop contribution
$V^\text{1-loop}$~\eqref{VMSSM1} vanishes at the vacuum.
In our example the gradient of the contribution $V^\text{1-loop}$ yields
three non-trivial conditions. We satisfy these gradient conditions
by an appropriate choice of the
renormalization scale~$Q$ and two additional parameters. In our example
we fix in this way besides $Q$ the generic scalar quark mass squared~$M_{\tilde{q}}^2$ as well as
the scalar top quark mass squared~$M_{\tilde{t}}^2$.

Explicitly, for the parameters given above we obtain
from~\eqref{vanishgradientK} a renormalization scale squared
\begin{equation}
\label{Qqfix}
Q^2 = 5.598 \cdot 10^6~\text{GeV}^2
\end{equation}
accompanied by $M_{\tilde{q}}^2= 42645~\text{ GeV}^2$ and
$M_{\tilde{t}}^2= 1.625 \cdot 10^7 \text{ GeV}^2$.
Following the approach outlined here we keep in this way 
the stationary point passing to the
effective potential. However, there could 
appear additional stationary points at the loop level. 
Here we will illustrate two strategies to verify that
the global minimum of the tree-level potential
is also the vacuum of the effective potential. 

The first strategy is to look for a deeper minimum of the
effective potential with numerical methods. With the known
vacuum at tree level and in addition with a fixed vanishing
gradient of the loop contributions via~\eqref{vanishgradientK}
we have a starting point for this search.
In this study we employ the minimization routine called
{\em Constrained Optimization by Linear Approximations}, (COBYLA)~\cite{Cobyla}.
This numerical minimization method is available in form of 
a C-library {\em Nlopt}~\cite{Nlopt}.
We can therefore confirm that the vacuum persists for the
case of the effective potential, since no deeper minimum is
found starting from this vacuum.  \\

In a second strategy we approximate the effective potential~$V^\text{1-loop}$
in a polynomial form. 
Explicitly we write the traces appearing on the r.h.s of~\eqref{VMSSM1}
as
\begin{equation}
\label{lnapprox}
\tr \bigg[ M^4 \bigg( \ln\bigg(\frac{M^2}{Q^2} \bigg) - \frac{3}{2} \big)  \bigg]
\approx
\frac{1}{Q^2} \tr \big( M^6 \big) - \frac{5}{2} \tr \big( M^4 \big)\;,
\end{equation}
with $M^2$ corresponding to a generic squared mass matrix.
The advantage of the polynomial form of the potential in this
approximation is that we
may apply methods like
the Gr\"{o}bner basis approach or homotopy continuation and
therefore are able to find systematically all stationary points
of the effective potential. These powerfull approaches are restricted to
polynomial equations.
In order to keep the vacuum passing to the effective potential we have
to apply the conditions~\eqref{vanishgradientK}.
Consistently, we apply these conditions 
taking the non-transzendental approximation of the from~\eqref{lnapprox} into account.
We remark that the traces of the squared mass matrices~$M_q^2$ and $M_{\tilde{q}}^2$ 
in~\eqref{lnapprox} can easily be computed
with help of computer algebra systems.\\

In this approximation we find 16 complex solutions of the stationarity equations 
for the case of the effective potential.
For the chosen parameter set, 
there remain four viable stationary points, that is, real solutions
with $K_0 \ge 0$, as
presented in Tab.~\ref{effmin}. This list is ordered from above to below
with respect to
decreasing potential values. Comparing the results with respect to the tree-level potential
in Tab.~\ref{treemin} with the results with respect to the effective potential, Tab.~\ref{effmin},
we see that we get a richer structure
of stationary points compared to the tree-level result.
The first line in Tab.~\ref{effmin} shows a fully broken minimum, corresponding
to the highest potential value among the stationary points.
Below this potential value appears a partially broken stationary point. This point has
an unwanted vacuum-expectation value of $v=1.89\cdot 10^6$~GeV, far above
$244.11$~GeV. 
With a decreasing potential value we find one stationary point corresponding to an unbroken
electroweak symmetry. This point does not correspond to a vanishing
potential unlike in the tree-level case.
In the effective case this point
gets additional contributions for vanishing fields from the loop terms.
Eventually, we find, as expected from the
conditions~\eqref{vanishgradientK}, a stationary point passing
from the tree-level, that is,  with the same values for the bilinears. 
As intended, this follows from
the vanishing gradient of the loop contributions at this point.
Let us note that although we find, as desired, the tree-level
stationary point also for the effective potential,
this point corresponds to a different potential value as can
be seen by comparing Tabs.~\ref{treemin} and \ref{effmin}.
This comes from the fact that the conditions \eqref{vanishgradientK}
require a vanishing gradient and {\em not} a vanishing potential
at the vacuum.
The partially broken stationary point is the deepest available and therefore
is identified with the global minimum. 
Of course, the revealed structure of stationary points 
is based on the polynomial approximation of the loop contributions,
which is only exact in the vicinity of the global minimum.

\begin{table}
\begin{tabular}{l|.|.|r r r r|.}
type of minimum & \multicolumn{1}{c|}{$v$~[GeV]} & \multicolumn{1}{c|}{$\tan(\beta)$} & \multicolumn{1}{c}{$K_0 [\text{GeV}^2]$} & \multicolumn{1}{c}{$K_1 [\text{GeV}^2]$} & \multicolumn{1}{c}{$K_2 [\text{GeV}^2]$} & \multicolumn{1}{c|}{$K_3 [\text{GeV}^2]$}  & \multicolumn{1}{c}{potential $[\text{GeV}^4]$}\\
\hline
unbroken & \multicolumn{1}{c|}0 & \multicolumn{1}{c|}{--} & 0 & 0 & 0 & 0  & 0\\
partially & 244.11 & 9.68 & 29795 & 6095 & 0 & $-29165$ & -5.79293\cdot 10^{7}\\
\end{tabular}
\caption{\label{treemin}Minima of the tree-level potential $V^\text{tree}$~\eqref{Vtree} with respect to their different types.
The first column gives the type of minimum with respect to the electroweak symmetry-breaking
behavior.}
\end{table}

\begin{table}
\begin{tabular}{l|.|.|r r r r|.}
type of minimum & \multicolumn{1}{c|}{$v$~[GeV]} & \multicolumn{1}{c|}{$\tan(\beta)$} & \multicolumn{1}{c}{$K_0 [\text{GeV}^2]$} & \multicolumn{1}{c}{$K_1 [\text{GeV}^2]$} & \multicolumn{1}{c}{$K_2 [\text{GeV}^2]$} & \multicolumn{1}{c|}{$K_3 [\text{GeV}^2]$}  & \multicolumn{1}{c}{potential $[\text{GeV}^4]$}\\
\hline
fully & \multicolumn{1}{c|}{--} & \multicolumn{1}{c|}{--} & 32632 &  -820010 & 0 & -32059 & 6.92704\cdot10^{11}\\
partially & 1.89 \cdot10^6 & 1.4\cdot10^{-6} & $1.79\cdot10^{12}$ & $-5.1\cdot10^{6}$ & 0 & $1.79\cdot10^{12}$ & 1.894\cdot 10^{6}\\
unbroken & \multicolumn{1}{c|}0 & \multicolumn{1}{c|}{--} & 0 & 0 & 0 & 0  & -1.76818\cdot10^{12}\\
partially & 244.11 & 9.6735 & 29795 & 6095 & 0 & $-29165$ & -1.76821\cdot 10^{12}
\end{tabular}
\caption{\label{effmin}Minima of the one-loop effective potential $V$~\eqref{effpot} with respect to their different types.
The first column gives the type of minimum with respect to the electroweak symmetry-breaking
behavior.}
\end{table}

Altogether, we can confirm the persistence of the vacuum passing
to the effective potential.
Firstly, numerical methods starting at the stationary point which is kept in the
effective potential does not find a deeper solution. Secondly, 
a systematic approach, looking for all stationary solutions in an
polynomial approximation of the loop contributions, does also not find
any deviation from the tree-level vacuum.
Nevertheless, we emphasize that
the loop contributions in general drastically change the potential and
therefore have to be taken into account in general. 
This was pointed out in
previous studies, for instance in~\cite{Gamberini:1989jw,de Carlos:1993yy,Casas:1995pd}.
In the example studied here we
detect two complex stationary solutions in the tree-level case whereas we
find 16 complex solutions in the (approximated) effective case. 
The two tree-level solutions, respectively
four solutions in the effective case pass the conditions that the bilinears
are real and $K_0 \ge 0$.

Note, that, for simplicity, we have only
considered the dominant loop contributions here. However the remaining
contributions may be implementes in a completely analogous way.

Let us finally comment on the chosen renormalization scale~$Q$.
First we recall that although the full loop expansion does not
depend on the renormalization scale any truncation in general does.
Of course, the value for the renormalization scale is an arbitrary mass scale.

Any physical Higgs potential has to provide a 
stable global minimum and therefore, a renormalization scale~$Q$
fulfilling~\eqref{vanishgradient},  or equivalently \eqref{vanishgradientK},
appears in a natural way. However, in models with anomalous, that is,
radiative electroweak symmetry breaking, there may be no viable global minimum
of the tree-level potential available which could pass to the
effective potential. In this case the
method proposed here is not aplicable.

In a supersymmetric model (with not too large soft-breaking terms),
the superpartners appear with opposite signs
in the Coleman-Weinberg contributions and we expect
the renormalization scale~$Q$ in the range of the superpartner masses.
In our example studied here this is indeed the case.
We have also compared the scale with a method to fix the renormalization scale
following from 
a vanishing grandient of the potential with respect to~$Q$; see for instance 
\cite{Gamberini:1989jw,de Carlos:1993yy,Casas:1994us,Casas:1995pd}.
We find that the scale choices agree within a factor of two.

Studying the two-loop effective potential with respect to minimization would
be very interesting, but is beyond the scope of this paper and is left for
future work. However, it is clear that
a scale~$Q$, fulfilling an analogous condition as in~\eqref{vanishgradient}
or \eqref{vanishgradientK},
has in general to be available for stability reasons.
Eventually let us remark that with increasing loop order it is to be expected that the dependence of
the vacuum-expectation values on the renormalization scale decreases.

\section*{Conclusion}

The minimization of Higgs potentials is in general a non-trivial task.
A thorough and straightforward way is to solve the
stationarity equations of the potential.
Proposed methods like the Gr\"obner-basis approach or
homotopy continuation are only applicable to polynomial systems of
equations. However, taking the Coleman-Weinberg loop contributions
into account the stationarity equations are no longer of polynomial form.
We have presented a method to nevertheless preserve the global minimum
of the potential considering the loop contributions.
The idea is to fix the renormalization scale, in addition to
other model scales, such that the gradient of the loop contributions
vanishes at the vacuum. We have verified in an explicit example that the minimum
of the effective potential appears to be valid, once
with numerical methods starting from the tree-level potential,
and once in a polynomial approximation of
the loop contributions. We have illustrated
the method for the case of the minimal supersymmetric model~(MSSM).
In particular, it could be confirmed that the structure of minima is
substantially changed passing from the tree-level to the quantum level
of the potential.

\begin{acknowledgments}
We would like to thank Joao~P.~Silva as well as
the unknown referee
for valuable comments.
D.M. acknowledges support by the U.S. Department of Energy under contract DE-FG02-85ER40237
and a DARPA Young Faculty Award.
\end{acknowledgments}

\bibliographystyle{h-physrev}

\end{document}